\documentclass[aps,prl,twocolumn,showpacs,superscriptaddress,preprintnumbers,amsmath,amssymb]{revtex4-1}

\usepackage{graphicx}
\usepackage{bm}
\usepackage{upgreek}

\begin{document}


\title{Controllable generation of a spin-triplet supercurrent in a Josephson spin-valve}

\author{Adrian Iovan, Taras Golod, and Vladimir M. Krasnov }
\email[E-mail: ]{Vladimir.Krasnov@fysik.su.se}


\affiliation{Department of Physics, Stockholm
University, AlbaNova University Center, SE-10691 Stockholm,
Sweden}

\begin{abstract}
It has been predicted theoretically that an unconventional
odd-frequency spin-triplet component of superconducting order
parameter can be induced in multilayered ferromagnetic structures
with non-collinear magnetization. In this work we study
experimentally nano-scale devices, in which a ferromagnetic spin
valve is embedded into a Josephson junction. We demonstrate two
ways of in-situ analysis of such Josephson spin valves: via
magnetoresistance measurements and via in-situ magnetometry based
on flux quantization in the junction. We observe that supercurrent
through the device depends on the relative orientation of
magnetization of the two ferromagnetic layers and is enhanced in
the non-collinear state of the spin valve. This provides a direct
prove of controllable generation of the spin-triplet
superconducting component in a ferromagnet.

\end{abstract}



\maketitle

An interplay of superconductivity (S) and ferromagnetism (F) in
hybrid S/F heterostructures leads to a variety of unusual physical
phenomena
\cite{Buzdin2005,Efetov2005,Fominov_2003,Blanter2004,Eschrig,Houzet_2007,Golubov,Trifunovic_2011,Melnikov_2012,Pugach_2012}.
Of particular interest is a possibility of generation of an
unconventional odd-frequency spin-triplet component of the
superconducting condensate \cite{Efetov2005,Golubov}. The
ferromagnetic exchange energy is usually much larger than the
superconducting energy gap. Consequently, a conventional
spin-singlet superconducting order parameter decays at a short
range $\sim 1$ nm in a spatially uniform, mono-domain ferromagnet.
Experimental observations of a long-range proximity effect through
strong ferromagnets \cite{Petrashov_1999,Wang_2010} and, in
particular, through almost fully spin-polarized half-metals
\cite{Pena_2004,Keizer_2006,Golod_2013} is consistent with
appearance of the spin-triplet component, which is insensitive to
strong magnetic and exchange fields. However, it may also be due
to various types of artifacts and, at certain circumstances, a
long-range spin-singlet component can be realized in clean S/F
heterostructures \cite{Melnikov_2012}. Therefore, unambiguous
confirmation for existence of the spin-triplet superconductivity
in S/F heterostructures requires controllable tunability of the
phenomenon. This is also prerequisite for potential applications
of S/F heterostructures in spintronics.

The spin-triplet order parameter in S/F heterostructures is
generated in presence of an active spin-mixing interface
\cite{Eschrig,Golubov} or in case of a spatially non-uniform
distribution of magnetization \cite{Efetov2005}. The latter can be
achieved in spin valve structures with several F-layers
\cite{Buzdin2005,Fominov_2003,Houzet_2007,Trifunovic_2011,Melnikov_2012,Pugach_2012}.
Both the spin-singlet and the spin-triplet components depend on
the angle between magnetization of F-layers in such
superconducting spin valves. The spin-singlet component is at
maximum for the antiparallel (AP) and minimum at the parallel (P)
state of the spin valve \cite{Melnikov_2012}. The spin-triplet
component is maximum at the non-collinear state with $90^\circ$
misalignment between magnetic moments and zero both in P- and
AP-states \cite{Fominov_2003,Trifunovic_2011}. Such a behavior has
been confirmed by analysis of the inverse proximity effect (i.e.,
suppression of superconductivity in an S-layer in contact with a
ferromagnet) for F/S/F \cite{Li_2013,Banerjee_2014} and S/F/F
\cite{Leksin_2011} structures.

Direct probing of the spin-triplet supercurrent in F-layers
requires measurements of perpendicular transport properties
through S/F heterostructures
\cite{Houzet_2007,Trifunovic_2011,Melnikov_2012,Pugach_2012}. Even
though supercurrent in such heterostructures has been observed
\cite{Bell_2004,Khaire_2010,Robinson_2010}, a conclusive evidence
for the spin-triplet nature of the supercurrent is still missing
due to a difficulty with separation of singlet and triplet
components and due to a general complexity of such a device with
several degrees of freedom, influence of stray fields and
Josephson vortices. Interpretation of the data becomes
particularly difficult in case of multi-domain switching of the
spin valve \cite{Khaire_2010,Blanter2004}. Consequently, for
unambiguous interpretation of the data it is necessary to study
small mono-domain structures and to establish accurate in-situ
characterization techniques.

Here we study nano-scale Josephson spin-valve devices, in which a
spin valve is implemented as a barrier in a Josephson junction. We
describe two methods for in-situ characterization of devices
using: (i) perpendicular magnetoresistance and (ii) in-situ
magnetometry based on flux quantization in a Josephson junction.
This way we unambiguously prove that the critical current is
enhanced in the non-collinear state of the spin valve,
successfully demonstrating a controllable generation of the
spin-triplet order parameter.

\begin{figure*}[t]
    \centering
    \includegraphics[width=0.9\textwidth]{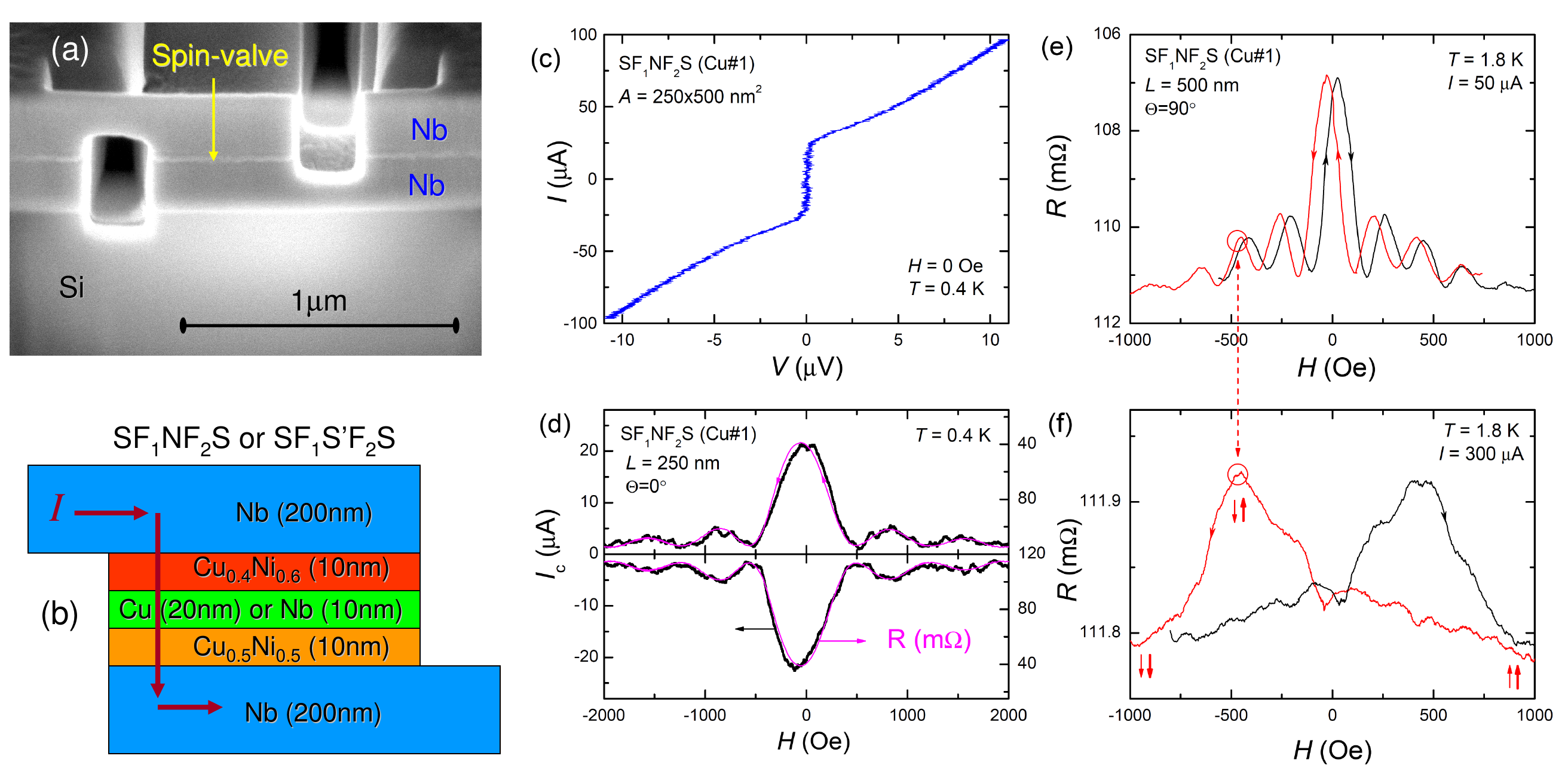}
    \caption{{\bf{Characterization of Josephson spin valves.}}
{\bf{a,}} SEM image of a device. {\bf{b,}} A sketch of studied
structures. {\bf{c,}} Current-Voltage characteristics of a
SF$_1$NF$_2$S junction (Cu$\#$1) at zero field. {\bf{d,}}
Fraunhofer modulation of the Josephson current (black symbols) and
a low-bias resistance (magenta lines) at $T=0.4$ K. {\bf{e,}} Low
and {\bf{f,}} high bias resistance of the same junction versus
magnetic field 
for upwards (black) and downwards (red lines) field sweeps at
$T=1.8$ K. It is seen that we can study both the Josephson current
and the spin-valve magnetoresistance by changing the bias level. A
hysteretic behavior of the spin valve is clearly seen in both
plots. Circles indicate the AP-state for the downward field sweep.
}
    \label{fig:fig1}
\end{figure*}

We study two types of Josephson spin-valves, consisting of two
dissimilar CuNi ferromagnetic layers $F_{1,2}$ separated by a
spacer layer of either a normal metal (N) Cu or a thin
superconductor (S') Nb. Scanning electron microscope (SEM) image
and a sketch of the structures are shown in Figures \ref{fig:fig1}
(a) and (b). The two ferromagnetic layers are made dissimilar in
order to achieve different coercive fields, required for
controllable switching of magnetization in the spin valve. This is
also necessary for generation of spin-triplet component of the
supercurrent. In the symmetric SFFS Josephson spin valve the
spin-triplet component cancels out, but in dissimilar SF$_1$F$_2$S
junction it does remain finite \cite{Trifunovic_2011}.

The SF$_1$NF$_2$S
(Nb/Cu$_{0.5}$Ni$_{0.5}$/Cu/Cu$_{0.4}$Ni$_{0.6}$/Nb
200/10/20/10/200 nm) and SF$_1$S'F$_2$S
(Nb/Cu$_{0.5}$Ni$_{0.5}$/Nb/Cu$_{0.4}$Ni$_{0.6}$/Nb
200/10/10/10/200 nm) multilayers were deposited by DC-magnetron
sputtering in a single deposition cycle without breaking vacuum.
The Cu$_{1-x}$Ni$_x$ films were deposited by co-sputtering from Cu
and Ni targets. 
Nano-scale junctions with sizes down to 100 nm were patterned by
photolithography, reactive ion etching and three-dimensional
nano-sculpturing using focused ion beam, as described in Ref.
\cite{Golod_2010}. Small dimensions were necessary both for
mono-domain switching of spin valves (domain size in CuNi is $\sim
100$ nm \cite{Veshchunov_2008}) and for enhancement of junction
resistances to comfortably measurable values. Measurements were
done either in a He-3 cryostat or in a He-4 gas flow cryostat. We
define the angle $\Theta=0$ and $90^{\circ}$ when the magnetic
field is applied along and perpendicular to the long side of the
junction, respectively. In all cases the magnetic field is
parallel to the junction plane. In total more than ten devices
were studied. The data below is representative for all of them.

\begin{figure*}[t]
    \centering
    \includegraphics[width=0.9\textwidth]{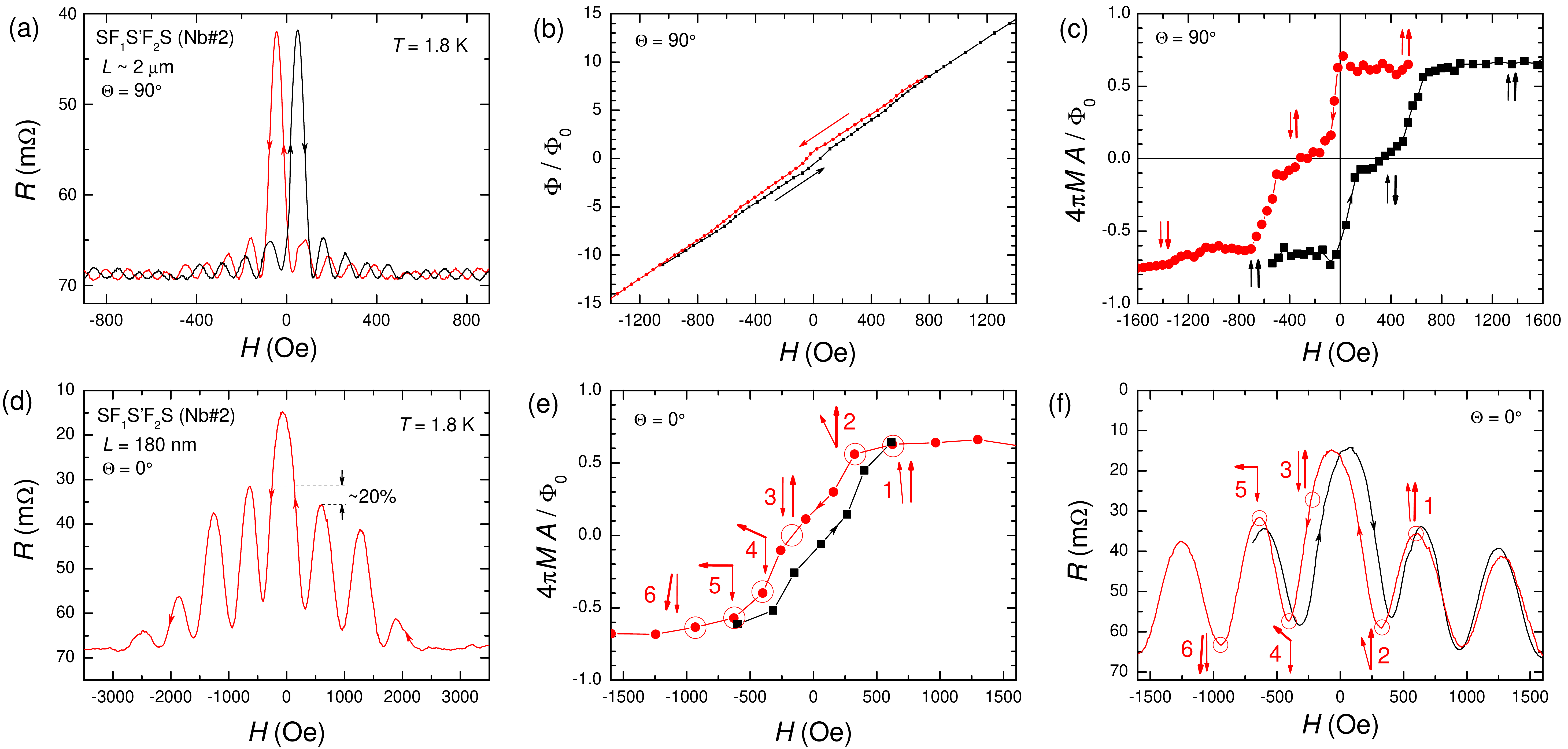}
    \caption{{\bf{Properties of SF$_1$S'F$_2$S junction: in-situ magnetometry and asymmetry of the critical current. }}
{\bf{a}}, Fraunhofer modulation of $R(H)$ at $\Theta=90^{\circ}$
for upward (black) and downward (red line) field sweeps. {\bf{b}},
Magnetic field dependencies of the flux in the junction. Each
point represents integer or half-integer $\Phi_0$, corresponding
to maxima or minima in $R(H)$ from panel (a). {\bf{c}},
Magnetization curves at $\Theta=90^{\circ}$, obtained from the
data in panel (b). The intermediate step with $M\sim 0$
corresponds to the AP-state of the spin valve. {\bf{d}},
Fraunhofer modulation of $R(H)$ at $\Theta=0^{\circ}$ for the
downward field sweep. {\bf{e}}, The magnetization curves at
$\Theta=0^{\circ}$. Arrows indicate orientations of magnetization
in the spin valve at points 1-6 for the downward field sweep.
{\bf{f}}, Central part of the Fraunhofer modulation of $R(H)$ at
$\Theta=0^{\circ}$ for upward (black) and downward (red line)
field sweeps. Note the asymmetry of the Josephson current at
points 1 and 5, corresponding to the same $|\Phi/\Phi_0|=1.5$. The
asymmetry is due to generation of an additional spin-triplet
component of the supercurrent in the non-collinear state of the
spin valve. }
    \label{fig:fig2}
\end{figure*}

Figure \ref{fig:fig1} (c) shows current-voltage ($I$-$V$)
characteristics of an SF$_1$NF$_2$S junction (Cu$\#1$ $\sim
250\times500$ nm$^2$) at $H=0$ and $T=0.4$ K. A critical current
$I_c \simeq 25~\mu$A is clearly seen. It corresponds to a critical
current density $J_c \simeq 2 \times 10^4$ A/cm$^2$. Black symbols
in Fig. \ref{fig:fig1} (d) represent magnetic field dependence of
the critical current at $\Theta=0^{\circ}$. The field is swept
from positive to negative values. A clear Fraunhofer-type $I_c(H)$
modulation proves Josephson nature of the supercurrent through a
spin valve, indicates good homogeneity of $I_c$ and a mono-domain
structure of F-layers \cite{Khaire_2010}. The supercurrent rapidly
decreases with increasing $T$ and becomes difficult to measure at
$T>2K$. To improve the resolution we performed lock-in
measurements of resistance with a small bias of the order of
$I_c$. The corresponding $R(H)$ modulation is shown by magenta
lines in Fig. \ref{fig:fig1} (d) (right axis). It is seen that
$I_c (H)$ is equivalent to the $R(H)$ data after appropriate
rescaling (reverse scale, large $R$ corresponds to small $I_c$).
Since the noise level is much smaller for lock-in measurements, in
what follows we will use low-bias resistance for characterization
of $I_c$.

Fig. \ref{fig:fig1} (e) shows the $R(H)$ modulation for the same
SF$_1$NF$_2$S junction at $\Theta=90^{\circ}$ and $T=1.8$ K.
Measurements were performed with a low ac-current amplitude
$I=50~\mu$A. Here we can clearly see a hysteresis between the
upward (black) and downward (red) field sweeps, which is due to
remanence magnetization of the spin valve. At higher fields (not
shown) Abrikosov vortices may be trapped in S-electrodes. As
discussed in Ref. \cite{Golod_2010}, vortex-induced hysteresis is
opposite to remanence magnetization and, therefore, can be clearly
distinguished. All the data presented here is for the vortex-free
case. The absence of vortices indicates that the magnetization
from F-layers do not puncture S-layers, but is forced to lie
in-plane despite possible perpendicular anisotropy of
magnetization in CuNi thin films \cite{Veshchunov_2008}.

Fig. \ref{fig:fig1} (f) shows the high bias resistance, measured
for the same configuration as in Fig. \ref{fig:fig1} (e) but with
a large ac-current $I=300~\mu$A $\gg I_c$. As seen from the
$I$-$V$ in Fig. \ref{fig:fig1} (c) in this case we measure
predominantly the normal resistance $R_n$ at the Ohmic part of the
$I$-$V$. It is seen that $R_n(H)$ represents a spin valve
magneto-resistance with minima and maxima at P- and
AP-orientations of magnetizations in the two ferromagnetic layers,
respectively \cite{Bell_2004,Dieny_1991}. From Figs.
\ref{fig:fig1} (e) and (f) it is seen that we can measure both the
critical current and the magnetoresistance by changing the bias
current level. Circles in Figs. \ref{fig:fig1} (e) and (f)
indicate the AP-state of the spin valve for the downward field
sweep. Thus we have successfully realized the Josephson
spin-valve, exhibiting both the spin valve effect and the
Josephson supercurrent.


Figure \ref{fig:fig2} represents data for an SF$_1$S'F$_2$S
junction (Nb$\#2$ $\sim 180$ nm $\times 2~\mu$m) at $T=1.8$ K.
Figs. \ref{fig:fig2} (a) and (d) represent $I_c(H)$ (low-bias
$R(H)$) modulations for magnetic field orientations perpendicular
to the long $\Theta=90^{\circ}$ and the short $\Theta=0^{\circ}$
sides of the junction, respectively. Minima and maxima of $I_c
(H)$ (maxima and minima of $R(H)$) correspond to integer and
half-integer flux quanta $\Phi_0$ within the junction. Due to a
significant difference in dimensions we see a significant
difference in flux-quantization fields for the two field
orientations. Unlike SF$_1$NF$_2$S junctions (Fig. \ref{fig:fig1}
(f)) the spin-valve magnetoresistance in SF$_1$S'F$_2$S junctions
is hardly detectable, probably due to much shorter scattering time
in Nb than in Cu. Therefore, we employ a different method for
determination of spin valve configuration in SF$_1$S'F$_2$S
junctions, following Ref. \cite{Bolginov_2012}, in which it was
demonstrated that flux quantization in a Josephson junction can be
used for in-situ analysis of magnetization.

In Fig. \ref{fig:fig2} (b) we plot the flux through the junction
as a function of applied magnetic field for the data from Fig.
\ref{fig:fig2} (a). Here every point corresponds to a maximum or a
minimum of $R(H)$. Apparently it represents the $B(H)=H+4\pi M(H)$
curve integrated over the junction crossection area $A$. At high
fields, when both F-layers are saturated in the P-state, the
$B(H)$ becomes linear. Subtracting this linear dependence we can
obtain the magnetization curve $M(H)$. Thus our junctions operate
as in-situ magnetometers (absolute fluxometers) for our nano-scale
spin-valves.

Figs. \ref{fig:fig2} (c) and (e) show thus obtained magnetization
curves for the two field orientations. From Fig. \ref{fig:fig2}
(c) it is clearly seen that upon sweeping of magnetic field the
magnetization of the spin valve switches via two steps. This is a
standard behavior of a spin valve with different coercive fields
of the two layers \cite{Dieny_1991}. First at $H\sim 200$ Oe the
weakest F$_1$ and later at $H\sim 500$ Oe the strongest F$_2$
layer switches the direction of magnetization. At $200$ Oe
$<H<500$ Oe there is a plateau with $M \sim 0$. It represents the
AP-state of the spin valve, as indicated in the figure. In Fig.
\ref{fig:fig2} (e) the behavior is similar, even though the
plateau is less defined.

Red arrows in Fig. \ref{fig:fig2} (e) indicate the configuration
of magnetization of the two F layers for downward sweeping of the
field.
At a large positive field, point-1, the spin valve is close to the
up-up parallel state. At point-2 the weak layer is partly rotated
and the strong layer remains in the up state. At point-3 the
magnetization becomes close to zero, which implies that the weak
layer has accomplished the rotation and the spin valve has
switched into the AP-state. At larger negative fields points-4 and
5 the stronger layer starts to progressively rotate downwards and
at point-6 the spin valve is close to the down-down parallel
state. Thus we can trace the state of the spin valve from the
in-situ magnetization measurement. This completes characterization
of the spin valve in our junctions and we can now proceed to our
main topic - discussion of controllable realization of the
spin-triplet component of the supercurrent.

In Fig. \ref{fig:fig2} (f) we replot the central part of the
$I_c(H)$ (inverted $R(H)$) modulation at $\Theta=0^{\circ}$, in
which we marked positions and magnetization orientations for the
points 1-6 from Fig. \ref{fig:fig2} (e). It is seen that for the
downward field sweep (red line) the critical current at point 1,
which correspond to $\Phi \simeq 1.5 \Phi_0$, is smaller than at
point 5, which correspond to $\Phi \simeq -1.5 \Phi_0$. The
asymmetry is also seen for other maxima of $I_c$ at half-integer
$\Phi_0$ in Fig. \ref{fig:fig2} (d). For the downward field sweep
all the maxima of $I_c$ at negative fields are larger than the
corresponding maxima at positive fields with the same absolute
value of $\Phi/\Phi_0$. As a consequence of this asymmetry there
are four lobes at the negative side and only three lobs at a
positive side of the $I_c(H)$ modulation in Fig. \ref{fig:fig2}
(d). The asymmetry is reversed for the upward field sweep, shown
by the black line in Fig. \ref{fig:fig2} (f). For the
SF$_1$NF$_2$S junction (Cu$\#1$) the same type of asymmetry is
seen from Fig. \ref{fig:fig1} (e). The field sweep direction
dependent asymmetry of $I_c(H)$ was observed in all studied
Josephson spin valve structures and is our central observations.

The observed left-right asymmetry of $I_c(H)$ is different from
the in-built $I_c(H)$ asymmetry caused by inhomogeneity of
junction parameters \cite{Krasnov_1997}, which does not depend on
the direction of the field sweep. We emphasize that such the
asymmetry was not present in our SFS junctions made with the same
technique and with the same dimensions, but containing only one
F-layer (see e.g., Fig. 4 (b) from Ref. \cite{Golod_2010}).
Consequently, the asymmetry is not the property of the individual
F-layers, but is related to the history dependent orientation of
the spin valve.

It is important to note that points 1 and 5 in Fig. \ref{fig:fig2}
(f) correspond to exactly the same absolute value of the flux
$|\Phi/\Phi_0| \simeq 1.5$. Consequently, the asymmetry is
entirely due to a different orientation of magnetization in the
spin valve. As shown in Figs. \ref{fig:fig2} (e) and (f) at point
1 the spin valve is close to the P-state, while at point 5 it is
in the non-collinear angle state. From the theoretical analysis it
follows that the spin-triplet component of supercurrent has a
maximum in the non-collinear state of SF$_1$F$_2$S junction with
dissimilar ferromagnets \cite{Trifunovic_2011}. Therefore, the
observed direction-dependent asymmetry of the supercurrent is
consistent with a controllable generation of the spin-triplet
component in our Josephson spin valves. The magnitude of asymmetry
indicates that the amplitude of generated spin-triplet
supercurrent is rather small, in the range of 10-20$\%$ of the
main spin-singlet part of the supercurrent. This is expected
because in SF$_1$F$_2$S structures the spin-triplet supercurrent
is only due to dissimilarity of F$_{1,2}$ layers
\cite{Trifunovic_2011}, which is not large in our case. On the
other hand, the dominant singlet component is beneficial for our
analysis. Singlet and triplet components of the Josephson current
are harmonic and double-harmonic, correspondingly, with respect to
the Josephson phase difference \cite{Trifunovic_2011}. Therefore,
the dominant spin-singlet component enables a regular, periodic in
$\Phi_0$, Fraunhofer $I_c(\Phi)$ modulation and facilitates
accurate characterization of our spin-valves via in-situ
fluxometry, as shown in Fig. \ref{fig:fig2} (c).

To conclude, we have successfully fabricated SF$_1$NF$_2$S and
SF$_1$S'F$_2$S Josephson junctions with embedded nano-scale spin
valve structures. We demonstrated that such Josephson spin valves
exhibit both the supercurrent and the spin-valve
magnetoresistance, both of which depend on the relative
orientation of magnetization of the two ferromagnetic layers. Flux
quantization in such structures was employed for in-situ
measurement of magnetization of the spin valve. Our main result is
the observation of an asymmetry of the critical current
with respect to the direction of sweeping
of the magnetic field, which depends solely on the orientation of
the spin valve.
In the non-collinear state of the spin
valve we observed
an increase of the Josephson supercurrent, which we attributed
to controllable generation of the spin-triplet component of the
order parameter.

We are grateful to A. Rydh for assistance in experiment and Ya.
Fominov for a valuable discussion. Technical support from the Core
Facility in Nanotechnology at Stockholm University is gratefully
acknowledged.




\end{document}